%% file: aw-ew-arxiv.tex
\documentclass[11pt,twocolumn]{article}

\usepackage{times}

\usepackage[labelfont=bf]{caption}

\usepackage[top=0.5in,bottom=0.75in,left=0.5in,right=0.5in]{geometry}

\usepackage{url}

\usepackage{subfig}

\usepackage{graphicx}

\usepackage{amsmath}

\graphicspath{{figures/}{plots/}{.}}

\usepackage[numbers]{natbib}
\usepackage[bookmarksopen,bookmarksnumbered,colorlinks=true,allcolors=blue]{hyperref}

\usepackage{xcolor}

\usepackage[compact]{titlesec}

\usepackage{helvet}
\usepackage{sectsty}
\allsectionsfont{\sffamily}

\usepackage{mathptmx}   

\usepackage{flushend}

\usepackage{framed}

\usepackage[strict]{changepage}

\usepackage{subfig}

\usepackage{mdframed}

\usepackage{multibib}
\newcites{postcog}{Additional References}
\newcites{sm}{Additional References}

\begin{document}


\title{\sffamily Anywhere \& Everywhere: A Mobile, Immersive, and Ubiquitous Vision for Data Analytics}

\author{\sffamily Niklas Elmqvist\\ 
\scriptsize\sffamily Aarhus University, Aarhus, Denmark}

\date{\sffamily September 2023}

\maketitle

\begin{abstract}
    Data is collected everywhere in our increasingly instrumented world and people are increasingly wanting to access this data from anywhere in it.
    This kind of \textit{anywhere \& everywhere data} present new challenges and opportunities for data-driven sensemaking and decision-making that will require leveraging novel mobile, immersive, and ubiquitous technologies undergirded by recent advances in human cognition.
    In this paper, we examine these emerging forms of analytics that are transforming how data analysis will be conducted in the future: in an ecosystem of connected devices, interactive visualizations, and collaborating users with vast amounts of data and analytical mechanisms available at their fingertips.
\end{abstract}


\begin{figure*}[tbh]
    \centering
    \subfloat[Anywhere and everywhere data scenario.]{
        \includegraphics[height=3.55cm]{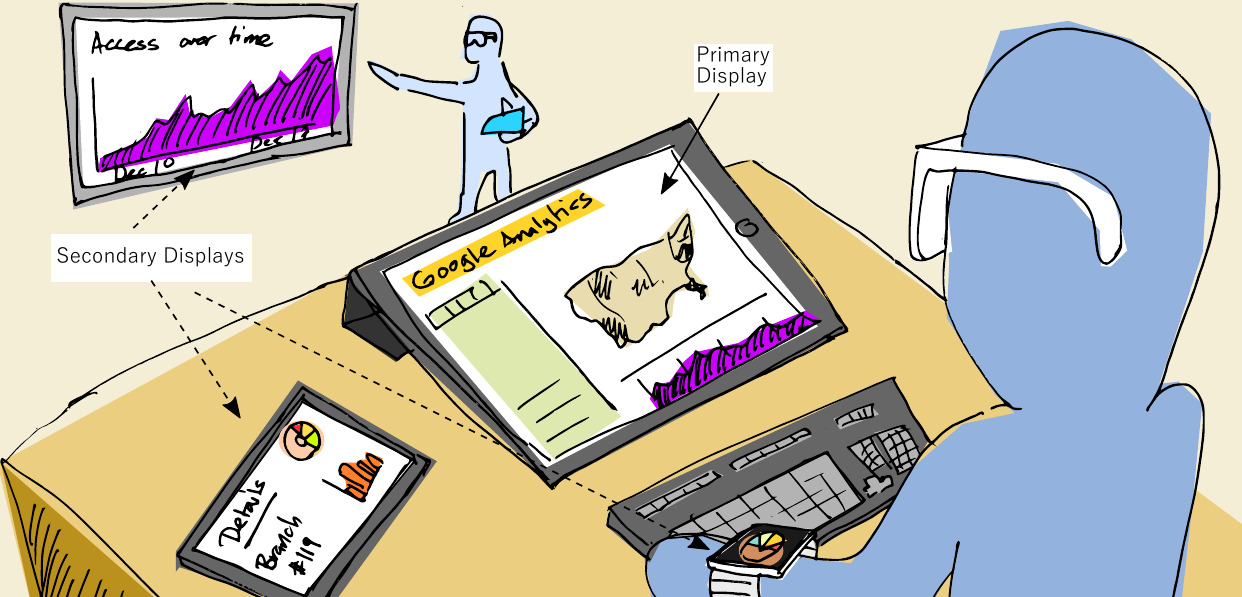}
    }
    \subfloat[Ubiquitous analytics using Augmented Reality.]{
        \includegraphics[height=3.55cm]{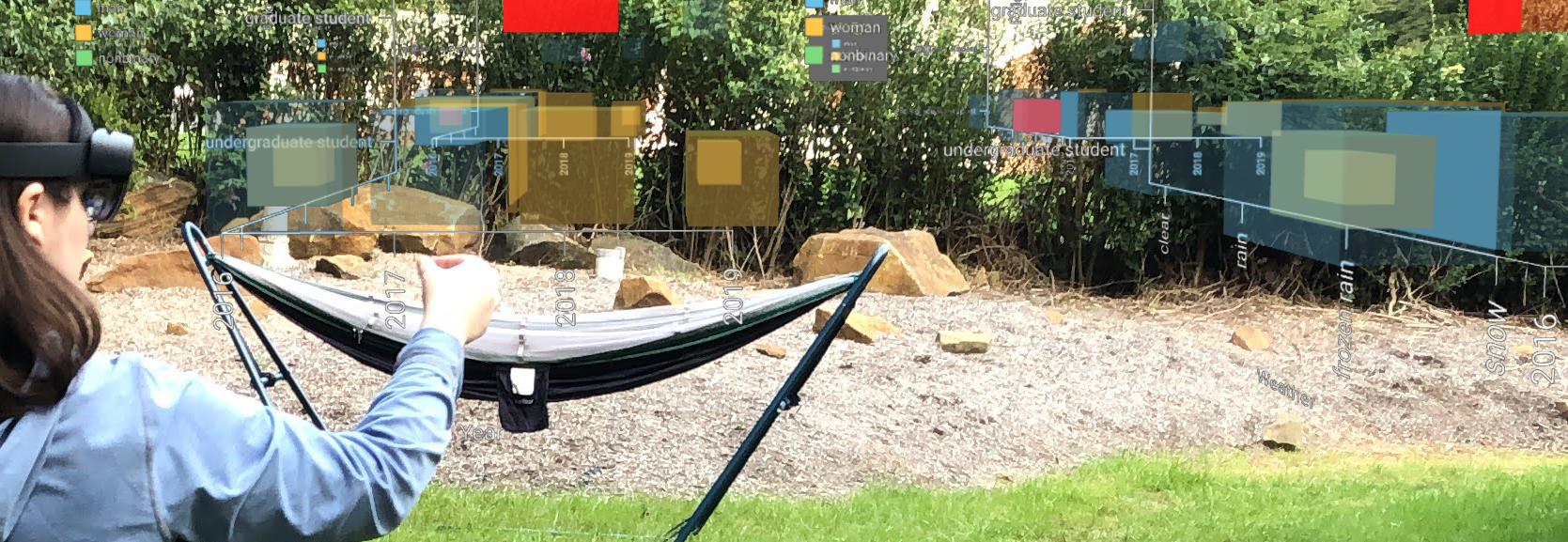}
    }
    \caption{\textbf{Mobile, immersive, and ubiquitous   sensemaking.}
    Two examples of data analytics conducted in a sensemaking environment of connected devices, interactive visualizations, and multiple collaborators.}
    \label{fig:teaser}
\end{figure*}

\input{content}

\section*{Acknowledgments}

The author thanks Zhicheng Liu for his careful feedback on this article.

\bibliographystyle{plainnat}
\bibliography{vis-stat}

\end{document}

%% file: content.tex
\section{Introduction}

Data is now collected everywhere and can be accessed anywhere.
Whether you are deciding which product to buy, which potential customer to visit, or just which lunch place to frequent, many situations in our everyday personal and professional lives benefit from access to relevant, accurate, and actionable data.
Such access support awareness, promote understanding, and help us make the right decisions in today's complex information society. 
This \textit{anywhere data} access is made possible by an increasing amount of \textit{everywhere data} that is collected from virtually all aspects of our physical and digital world~\cite{DBLP:journals/interactions/FisherDCD12}: shopping lists and purchase histories; movie, music, and book preferences; electronic health records and medical test results; colleague, friend, and family relationships; professional experience and education, and so on.
While significant privacy, security, and safety concerns are intrinsic to this confluence of anywhere and everywhere data, there is today also an unprecedented opportunity to use this data to support individuals in navigating the complexities of their professional and personal lives.
Fortunately, the last 20 years of the mobile revolution has given us the means to achieve this; mobile computing is now the dominant computing platform on the planet, with more than 15 billion mobile devices in existence in 2020\footnote{\url{https://www.statista.com/statistics/245501/multiple-mobile-device-ownership-worldwide/}} and more than 6 billion of them being ``smart'' and able to access the internet.\footnote{\url{https://www.statista.com/statistics/330695/number-of-smartphone-users-worldwide/}}
However, these devices---for all their mobility---currently are mere portholes into the digital world, are rarely designed to work together effectively to support one user, let alone multiple ones, and lack the powerful analytical tools needed to enable data-driven decision-making on the go.

Fortunately, this may be about to change, with mobile and ubiquitous computing~\cite{Weiser1991} as well as extended reality (XR)~\cite{speicher19whatismr} finally beginning to transform the fields of visualization and data science.
Applying these technologies to data analysis suggests a future of ubiquitous~\cite{Elmqvist2013} and immersive analytics~\cite{Marriott2018} where clusters of networked mobile devices form an ecosystem for data analytics that can be accessed anytime and anywhere.
Such a vision of mobile, immersive, and ubiquitous sensemaking environments would blend state-of-the-art analytics methods with our physical reality to enable making sense of any kind of data in virtually any situation.

However, we shouldn't be weaving computation into our everyday lives just because we \textit{can}.
Rather, progress in cognitive science~\cite{Clark1998, Hutchins1995, Shapiro2011} supports leveraging the new generation of mobile, immersive, and ubiquitous technologies towards data analytics.
These so-called \textit{post-cognitive frameworks} suggest that human thought is not contained merely within our heads, but encompasses the entire ecosystem~\cite{Liu2008} of other people, physical artifacts in our surroundings, and our very own bodies.
Distributing computational nodes into our physical surroundings will thus enable us to better scaffold analytical reasoning, creativity, and decision making.

In this paper, we will investigate how the prevalence of collected everywhere data can enable leveraging it for anywhere and anytime access.
To achieve this, we will first explore the concepts of anywhere and everywhere data, and see how current technologies can (and cannot) support cognition in these ubiquitous computing environments.
We will also review current research in ubiquitous and immersive analytics that build toward this vision.
Finally, we synthesize the current research challenges facing the scientific community and describe the outlook for future research on the topic.

\begin{figure}
    \centering
    \includegraphics[width=\linewidth]{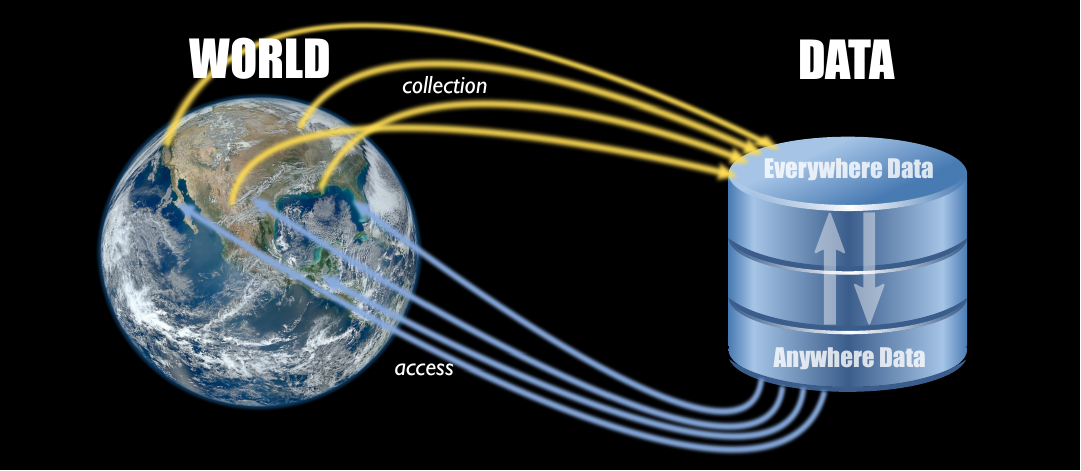}
    \caption{\textbf{Anywhere and everywhere data.}
    Today data is collected from everywhere and is increasingly being accessed from anywhere.}
    \label{fig:anywhere-everywhere}
\end{figure}

\section{Anywhere \& Everywhere Data}

The ``third wave'' of computing---ubiquitous computing---envisioned by Mark Weiser and Xerox PARC in the late 1980s~\cite{Weiser1991} is essentially here, even if it looks subtly different from their original vision~\cite{Dourish2011}.
Instead of talking alarm clocks, we have voice-driven home assistants; instead of cheap and disposable computational ``tabs,'' we have smartphones that we bring everywhere we go; instead of an interactive liveboard in every office, we have Zoom and Google Meet videoconferences at our beck and call.
Nevertheless, with billions of mobile devices in existence in the world today---many of them smart---and a burgeoning Internet of Things making increasing inroads into our physical reality, it is safe to say that we are rapidly approaching a world where computing has indeed been woven into everyday life.
The critical difference is that, unlike in the original vision of ubiquitous computing, the implementation has been more about personal devices than shared infrastructure---what Harrison et al.~\cite{DBLP:journals/ieeemm/HarrisonWD10} call \textit{quality} rather than \textit{quantity} computing.

Regardless, this quiet ubiquitous and mobile computing revolution has had two very specific outcomes relevant to data analytics (Figure~\ref{fig:anywhere-everywhere}): the emergence of (1) everywhere and (2) anywhere data.
For the former, it has led to digital data being collected across all of our society and in virtually all walks of life, both in the real world and online.
The Internet of Things encompasses a wide and growing array of devices, such as webcams and security cameras, smart thermostats and light bulbs, digital weather stations, pollution and air quality sensors, smart locks, connected home appliances, etc.
These devices are also increasingly becoming entwined into professional settings, such as for connected healthcare devices and monitors, autonomous farming equipment, wireless inventory trackers, biometric security scanners, wireless sensor networks, and unmanned military equipment.
Common for all of the datasets collected by these devices is that they are local, temporal, and contextual:

\begin{itemize}
    \item\textbf{Local:} connected to a specific geographic or semantic location (e.g., the temperature on top of the Eiffel tower);
    \item\textbf{Temporal:} associated with a point in time or a temporal pattern (e.g., the temperature on May 1); and
    \item\textbf{Contextual:} related to a specific situation, and thus best interpreted in that context (e.g., a car engine's temperature when traveling at 60 miles per hour).
\end{itemize}

The situation is slightly different for data collected online, as there may be no geographical location associated with the data. 
Furthermore, in many cases, even data collected from the real world is stored in databases where these local, temporal, and contextual aspects are discarded or aggregated.
For example, the thousands of webcams located around the world that merely show video to whoever happens to tune in represent a lost opportunity; what if we instead continuously ran privacy-preserving image analysis algorithms on this footage to capture environmental data such as wind speed, people density, snow accumulation, beach erosion, or road traffic?
Nevertheless, it is safe to say that virtually everything that we do online is tracked and recorded, a tasks that is only made easier by the fact that in the digital world, instrumentation is trivial.

The second outcome from the mobile computing revolution is that technology advances has instilled in users a desire to be able to access this data anytime and anywhere.
Many people are accustomed to enjoying near-constant connectivity with the internet and all of its trappings, from social media and video streaming to email and instant messaging.
The step to expecting the same anywhere data access even for analytics and sensemaking tasks is not far. 
However, while current devices certainly are capable of accessing, managing, and storing these datasets, input and output technology has only recently reached a level where such sensemaking can be conducted.
But \textit{how} should we go about doing so effectively?

\smallskip

\begin{mdframed}[backgroundcolor=orange!10,
    frametitle={\section*{Post-Cognitive Frameworks}},
    frametitlerule=true, frametitlebackgroundcolor=brown!50]

Traditional cognitive science views the individual human as a fundamental unit of cognition, with a basic workflow consisting of perception to understand your surroundings, cognition within the mind, and physical action to manipulate the surrounding environment.
It follows that in traditional cognitive science, the mind maintains a model of the world, and the senses are used to continuously update this model.
Such a model is called \textit{solipsism}, and holds that you can only be sure about your own mind and not the external world; in essence, you could replace the world with simulated sensor input and muscle output---as in the movie \textit{The Matrix} (1999)---and you would be none the wiser.
However, the last few decades of research have slowly chipped away at this solipsistic view of the mind.
For example, post-it notes, notebooks, and smartphones trivially extend our memory~\cite{Clark1998}.
Discussions with other people facilitate thinking and their abilities complement ours, forming a socially distributed cognitive system~\cite{Hutchins1995}.
Kirsh and Maglio~\citepostcog{DBLP:journals/cogsci/KirshM94} found evidence that expert Tetris players perform costly geometric rotation in the world rather in their minds. 
In other words, cognition is essentially a product of an individual's interaction with their environment rather than a closed system inside the individual's mind~\cite{Liu2008}.

Frameworks that go beyond the classic human information processing model are called \textit{post-cognitive frameworks}.
We will discuss three particular ones here, all of which have useful properties that we draw upon in this paper.
However, we note that these frameworks, as well as cognitive science as a whole, is still very much a work in progress and thus---to bastardize a phrase by statistician George E.P.\ Box---that they may all be incorrect to some degree, but some may still be useful.

\begin{itemize}
    \item\textbf{Extended cognition:} The unofficial motto of Clark and Chalmers' extended cognition framework is that ``Cognitive processes ain't (all) in the head!''~\cite[p.\ 8]{Clark1998}, arguing that the surrounding environment is a fundamental component of all human cognition---that, in fact, we're all effectively cyborgs, but in the most natural way in symbiosis with our environment.
    
    \item\textbf{Socially distributed cognition:} Developed by UCSD cognitive psychologist Ed Hutchins~\cite{Hutchins1995} in the 1980s and 90s, socially distributed cognition (DCog) holds that human cognition is distributed across sociocultural systems in the individual's surroundings, which includes physical artifacts, other individuals, and cultural systems (history, practice, etiquette, etc).
    
    \item\textbf{Embodied cognition:} The basic tenet of embodied cognition is that thinking is fundamentally influenced by and inseparable from our own bodies~\cite{Shapiro2011}.
    These ideas have been fundamental to the areas of embodied and tangible interaction in HCI~\citepostcog{Dourish2001}.
    
\end{itemize}

Beyond these three, we also note Scaife and Roger's concept of \textit{external cognition}~\citepostcog{Scaife1996} in support of graphical representations, which, even if it does not quite reach the level of a framework, discusses how visualization helps cognition by offloading computation and memory, re-representing data in a more suitable form, and graphically constraining inferences.

\bibliographystylepostcog{ACM-Reference-Format}
\bibliographypostcog{anywhere-everywhere}
\end{mdframed}

\section{The Cognitive Case}

Advances in cognitive science summarized as so-called \textit{post-cognitive frameworks}---such as embodied~\cite{Shapiro2011}, extended~\cite{Clark1998}, distributed cognition~\cite{Hutchins1995} (see sidebar)---suggest that human thinking is a system-level process~\cite{Liu2008} not merely contained within our brains, but which expands to also include the world around us, the physical artifacts in our vicinity, our own bodies, as well as other people.
In a post-cognitive framework, cognition is represented as information being transformed from one media to another through interactions, such as a person using a pen to write a reminder on a post-it note, placing the note on a refrigerator, and then reading the reminder at a later date and acting upon it.
In other words, tools do not amplify our mind, but instead transform certain cognitive activities---such as remembering long number sequences---into other, less taxing cognitive activities---such as reading.
Analogously, a visual representation of data on a digital screen is another form of media that a person can interact with in order to view, manipulate, and understand the underlying data---but so is the notepad the person uses to jot down notes, the phone through which they speak to a colleague, and the calculator they employ for quick arithmetic. 

Effectively supporting sensemaking thus means instrumenting the entire ecosystem of artifacts involved in the cognitive process.
This is in contrast to traditional human-computer interaction (HCI) and visualization paradigms, which tend to consider only the actual interface between user and machine: the visual output from the screen, and the user input from the mouse, keyboard, touchscreen, or microphone.
By distributing interactive representations of data on a multitude of digital devices scattered around our physical surroundings as well as with the collaborators involved in the cognitive system, we are for all intents and purposes creating a cybernetic extension of the mind, expanding it using these digital devices.
This notion would have been revolutionary if not for the fact that we have already been doing this for thousands of years ever since the first human picked up the first fallen branch and used it to dig a hole in the dirt.
Just like a shovel is a cybernetic extension of a person's arms to enable them to dig better, and an excavator takes this concept of human extension even further, so a computer can be seen as a cybernetic extension of our logic, reasoning, and memory---or, as Steve Jobs famously puts it, a ``bicycle for the mind.''
Nevertheless, from the digging example, it is clear that the type of tool makes a big difference, and even if digging may be a solved problem, the same cannot be said of sensemaking.

Of course, the argument here is not that the more devices we weave into our everyday environment, the more effective the cognitive process.
As Harrison et al.~\cite{DBLP:journals/ieeemm/HarrisonWD10} point out with their quality vs.\ quantity computing argument, human attention is a scarce commodity, and the current focus on a single, highly-capable personal device seems to have won out over the ``one user to many devices'' ubicomp vision from the 1990s.
However, there clearly exists a middle ground between a single device per person and dozens.
In addition, nothing says that the devices have to be homogeneous; it may be difficult to use two tablets at the same time, but a tablet and a smartwatch can be complementary. 
For one thing, data analysts do not only use computers but tend to surround themselves with pen and paper, calculators, reference material, books, and other physical artifacts, not to mention other analysts~\cite{Liu2008}.
For another, physical space is a key factor in the data analysis process; Wright et al.~\cite{Wright2006} describe how some professional intelligence analysts would use the entire floor of their office to arrange documents during analysis, and Andrews and North conducted empirical studies demonstrating the utility of significant visual space in facilitating sensemaking~\cite{Andrews2010}.
In fact, the intelligent use of space as part of cognition is a cornerstone of post-cognitive frameworks~\cite{DBLP:journals/ai/Kirsh95}, simplifying choice, facilitating perception, and aiding computation.
Furthermore, \textit{proxemics}---hailed by some as the new ubicomp~\cite{Greenberg2011}---tells us how people relate physically to artifacts and other people as they interact with them; for example, we typically turn to face people we speak to.
These are all prime arguments in favor of instrumenting all of the components involved in the cognitive process---ranging from notepads, books, and other people---using networked digital devices.
Drawing from all three of the ubiquitous computing, visual computing, and visual analytics traditions, we call this approach \textit{ubiquitous analytics}~\cite{Elmqvist2013}.

\section{The Gap in Our Tech}

If cognitive science supports the use of these networked groups of devices to facilitate sensemaking, how come we have not yet seen a plethora of such ubiquitous analytics systems on the market and in the scientific literature? 
The answer is that until only recently, the technology required to harness such anywhere and everywhere data has been outside our reach.
More specifically, the gaps in our technology include the screen size, the input surfaces, and the general design of the mobile devices we need for this endeavor.

Both screen size and input surfaces suffer from a device miniaturization tradeoff, where mobile devices already maximize the input and output dimensions to encompass the full size of the device itself.
They simply cannot be made much larger to avoid diminishing the mobility and portability of a smartwatch, tablet, or smartphone.
One solution for both input and output is to go beyond the device itself to appropriate the physical world as part of the interaction~\cite{Harrison2010}.
For example, portable projectors can turn any nearby surface into a display of arbitrary size~\cite{DBLP:journals/interactions/DachseltHJLRR12}.
Depth cameras, ultrasound, or electrical sensing can similarly transform walls, tables, or even our very own bodies into touch surfaces.
The rise of consumer-level mixed and augmented reality~\cite{speicher19whatismr} equipment has taken this idea even further by turning our entire world into a potential canvas for data display and manipulation.
Instead of our smartphones being mere portholes into an unseen world of data, XR technology has broken down the fourth wall hemming us in.

A hurdle remains, however: current mobile devices are still designed using the ``quality computing'' mindset.
This means that each individual device is intended to be used in isolation and with the undivided attention of the user.
A significant gap for mobile computing is to derive new design paradigms where multiple devices can stack together and scaffold each other for the current task while minimizing barriers and tedious housekeeping.
We will demonstrate some examples of how to design such stacking devices in the treatment to come. 

\smallskip

\begin{mdframed}[backgroundcolor=orange!10,
    frametitle={\section*{Sensemaking \& Visualization}},
    frametitlerule=true, frametitlebackgroundcolor=brown!50]

Sensemaking is the activity of searching for a representational schema to fit available data~\citesm{Klein2006, Russell1993} (e.g., a mental model), allowing people to attribute meaning to available data and answer questions about it.
In terms of the cognitive foundations discussed earlier in this paper, this activity often involves a combination of internal and external cognitive representations, such as data tables, charts, notes, sketches, and calculations.
These representations should ideally be chosen to facilitate mental activities required (memory, computation, planning), thereby minimizing the \textit{cost structure}~\citesm{Russell1993} of cognition.

Data visualizations have long been used for communicating insights because of their accessible and visual form.
However, visualizations are also effective for providing data-driven overviews, generating hypotheses, and answering questions at little cognitive cost.
This makes visualization particularly suitable for sensemaking, where exploring the data for meaning and hypotheses is a significant part.
Such \textit{exploratory data analysis}~\citesm{Tukey1977} represents an alternative to more traditional hypothesis-driven \textit{confirmatory data analysis}.

\bibliographystylesm{ACM-Reference-Format}
\bibliographysm{anywhere-everywhere}
\end{mdframed}

\section{Ubiquitous, Immersive \& Situated Analytics}
 
While command-line tools, automated scripts, and libraries are common in general data science tasks such as computation, wrangling, and confirmatory analysis~\cite{Cao2018}, the kind of exploratory data analysis common to sensemaking (see sidebar) often benefits from a more fluid and visual interaction model.
This is particularly true in mobile settings, where the precise text entry required for programming and command-line interfaces remains challenging on handheld devices.
In such situations, it is more convenient to turn to interactive visual interfaces and automated recommendations.
The scientific field of \textit{data visualization} concerns itself with precisely these visual and interactive data displays. 
In fact, the field has recently undergone a dramatic change with the rise of ubiquitous~\cite{Elmqvist2013}, immersive~\cite{Ens2021, Marriott2018}, and situated analytics~\cite{elsayed16situateddef} for tackling these use-cases.

Three-dimensional visual representations have long been the norm in many scientific applications for visualization, such as flow visualization, medical imaging, and volumetric rendering. 
However, the very first applications of data visualization to immersive settings actually came from outside the visualization field.
In 2003, Bowman et al.~\cite{DBLP:conf/vrst/BowmanNCPPY03} proposed a research agenda for so-called \textit{information-rich virtual environments} (IRVEs) that combined 3D Virtual Reality with information visualization. 
Touch-enabled tabletops were an early platform, with Spindler and Dachselt~\cite{Spindler2010} proposing a tangible lens for interacting with data on or above a horizontal display. 
Finally, visualization made inroads into mobile computing in a much more unobtrusive manner, with the first applications being commercial ones on smartphones.
Lee et al.~\cite{Lee2022} review mobile data visualization in a recent book on the topic.

Starting in 2011, I joined this research area by proposing the notion of an embodied form of human-data interaction~\cite{Elmqvist2011c}.
This launched my research agenda on this topic, and my students and I followed this up with an embodied lens used for exploring data on a touch-based tabletop~\cite{Kim2012}. 
It eventually led to Pourang Irani and I defining the \textit{ubiquitous analytics}~\cite{Elmqvist2013} paradigm in 2013, which serves as an umbrella term for the research field: anytime, anywhere sensemaking performed on a plethora of networked digital devices, not just immersive ones.
My students and I also proposed several computing infrastructures for realizing this vision for data analytics, including Munin (a peer-to-peer middleware based on Java)~\cite{Badam2015}, PolyChrome (which was the first web-based framework)~\cite{Badam2014a}, and Vistrates (a mature web replication framework for component-based visualization authoring)~\cite{Badam2019}.
Researchers have since taken our ideas further; for example, ImAxes~\cite{cordeil17imaxes} enables data analysis in Virtual Reality using powerful hand gestures for connecting dimensional axes in 3D.
Another early example was Butscher et al.'s work combining parallel coordinates display in augmented reality on top of a tabletop display~\cite{butscher2018clusters}.

Along the way, several variations of these ubiquitous forms of analytics have emerged. 
\textit{Immersive analytics}, initially introduced by Chandler et al.~\cite{Chandler2015} in 2015, take a specific focus on immersive and 3D spatial technologies to support sensemaking~\cite{Marriott2018}. 
\textit{Situated analytics}~\cite{elsayed16situateddef}, on the other hand, emphasizes the spatial referents for data in the real world.
Willet et al.~\cite{Willet2017_Embedded} built on this in 2017 for \textit{embedded data representations} that are deeply integrated with the spaces, objects, and contexts to which the data refers.

Given the one-to-many ratio of users to devices in traditional ubicomp, cardinality is a common denominator also for ubiquitous analytics.
In the below subsections, we discuss cardinality from three different perspectives and give specific examples for each: multiple devices, multiple resources, and multiple collaborators.

\begin{mdframed}[backgroundcolor=black!5, linewidth=0pt]
\textbf{Devices Now and in the Future:} In this vision of an increasingly ubiquitous and mobile future, the traditional view of a ``device'' as a physical artifact with computation, storage, memory, display, and input is being challenged.
Some such physical devices will surely remain, but in other ways, devices will recede into the background or be replaced by a single mixed/augmented reality display.
\end{mdframed}

\begin{figure}[htb]
    \centering
    \includegraphics[width=\linewidth]{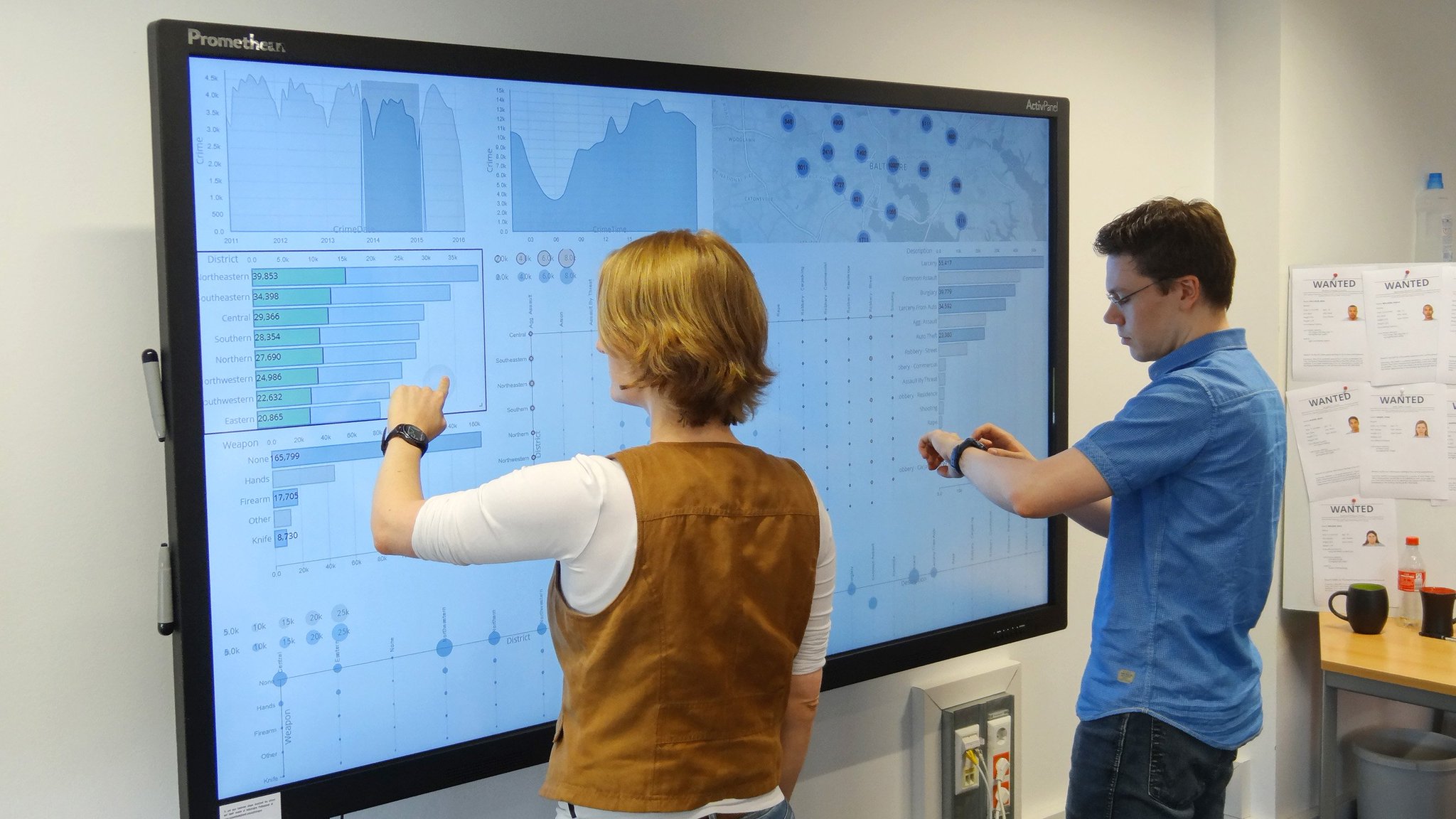}
    \caption{\textbf{Two users collaborating using the David and Goliath system.}
        The woman on the left is interacting with the large touch display (``Goliath''), which is---by its size and orientation---a public and shared surface.
        The man on the right is interacting with his smartwatch (``David''), which is used for personal storage and private data manipulation.
    }
    \label{fig:david-and-goliath}
\end{figure}

\begin{figure*}[tbh]
    \centering
    \includegraphics[width=\linewidth]{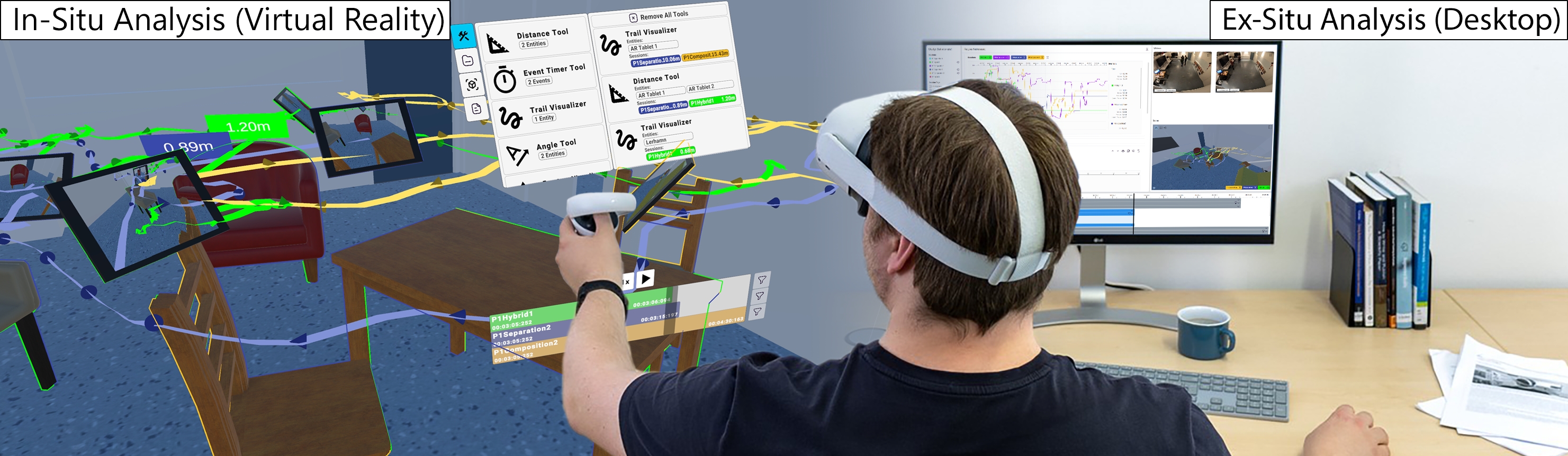}
    \caption{\textbf{Hybrid interfaces in ReLive.}
        The 3D VR interface (left) and the 2D desktop interface (right) are used depending on the task.
        First-person 3D invites situated analytics, whereas the 2D desktop better supports higher-level analysis.
    }
    \label{fig:relive}
\end{figure*}

\subsection{Multiple Devices}

The first factor worth investigating in a ubiquitous approach to analytics is how to best manage the multiple types of devices that a user will engage with during sensemaking. 
As already mentioned, human attention is a finite resource that must be managed judiciously.
Current device platforms are typically designed for focused use; for example, a smartphone engages the user's hand, often the dominant one, so adding a second smartphone is seldom helpful.
Rather, adding multiple devices to an interaction should be complementary, either by physical form factor (e.g., a smartwatch or a large display), by physical placement (e.g., hand-held, wall-mounted, or head-mounted), by task (e.g., a primary display used for map navigation and secondary ones to show legends and drill-down details), or by combinations of these.
Users are rarely helped by two identical devices inhabiting the same position in this design space, except possibly for comparison tasks where holding two tablets side by side may be beneficial.

We studied this phenomenon in a research paper from 2018 on the interplay between smartwatches and large touch displays~\cite{Horak2018} that we informally dubbed ``David \& Goliath.''
The form factor of these two different computing platforms are radically different: a smartwatch is a fundamentally personal device, whereas a large display is a fundamentally public one. 
Smartwatches have small displays, are attached to a person's wrist and are thus always within reach, and are only really accessible---both in terms of physical reach as well as social practices---to the wearers themselves.
Any action performed and any data displayed on a smartwatch will accordingly be personalized. 
In contrast, a large touch display is often vertically mounted on a wall, is visible to many by virtue of its size, and invites interaction by anyone within physical reach.
This, in other words, is a prime example of how a ubiquitous analytics environment can be designed to use complementary device platforms.

Figure~\ref{fig:david-and-goliath} shows an example of the David \& Goliath system in action. 
Drawing on the personal affordances of the smartwatch, we mainly use the smaller display as a personal storage container as well as a mediator and a remote control---all tasks that are well-suited to the device.
The large display, on the other hand, is a public and shared display that provides the fundamental data visualization for the team working on a specific task.
Because of its size, it can show dashboards of multiple visualizations rather than just one.

Devices can also complement each other over time. 
In the recent ReLive~\cite{DBLP:conf/chi/HubenschmidWFBZ22} system (Figure~\ref{fig:relive}), we proposed a hybrid desktop/VR interface that gives the same user access to two different ubiquitous analytics interfaces depending on task. 
ReLive was built to help mixed reality designers and researchers analyze real-time telemetry from user studies conducted in virtual or augmented reality environments.
Thus, the primary data being visualized is 3D tracking and event logs over time.
Here, again, the two analysis interfaces---an in-situ 3D Virtual Reality interface and an ex-situ 2D desktop interface---are complementary in that each is suited for a specific task.
When it comes to understanding 3D data, such as how two participants worked together in a 3D assembly task, nothing beats viewing the data from a first-hand perspective in 3D. 
However, when deriving high-level abstract findings, such as the average distance between participants over an entire session, the 2D desktop interface is optimal.
Having access to both interfaces allows a user to pick the right tool depending on task.

But it is not enough that two devices are complementary---they should actively support each other.
This is particularly important in ReLive, where the same user will be switching between the two different interfaces time and time again, sometimes as part of the same task.
This interface switch leads to a ``cognitive context switch,'' where the user must reorient themselves within the new interface in relation to the other interface.
In ReLive, this is facilitated by providing common anchors between the two views.
The 2D desktop view provides a 3D in-situ viewport that is visible at all times.
Analogously, the 3D Virtual Reality view replicates many of the interface features from the desktop view, and selections made inside the first-hand perspective can be accessed in the 2D view.

\begin{mdframed}[backgroundcolor=black!5, linewidth=0pt]
\textbf{Takeaway:} Each device involved in ubiquitous analytics should have a clear role that is complementary to the other devices and transitions between them should be seamless.
\end{mdframed}

\subsection{Multiple Resources}

Given that the above discussion is at the granularity of entire devices, then our next concern is how to manage the resources associated with these devices, such as displays, computational power, input affordances, etc. 
Resource management quickly becomes onerous and almost all-consuming when multiple devices are involved, and automation is thus required.
For example, merely imagine logging into your typical workplace network, and then multiply this for every device you want to include in a work session.
If the overhead associated with using multiple devices becomes overwhelming, people will simply stick to a single device.

\begin{figure*}
    \centering
    \subfloat[Ubiquitous analytics scenario involving multiple devices.]{
        \includegraphics[height=4.5cm]{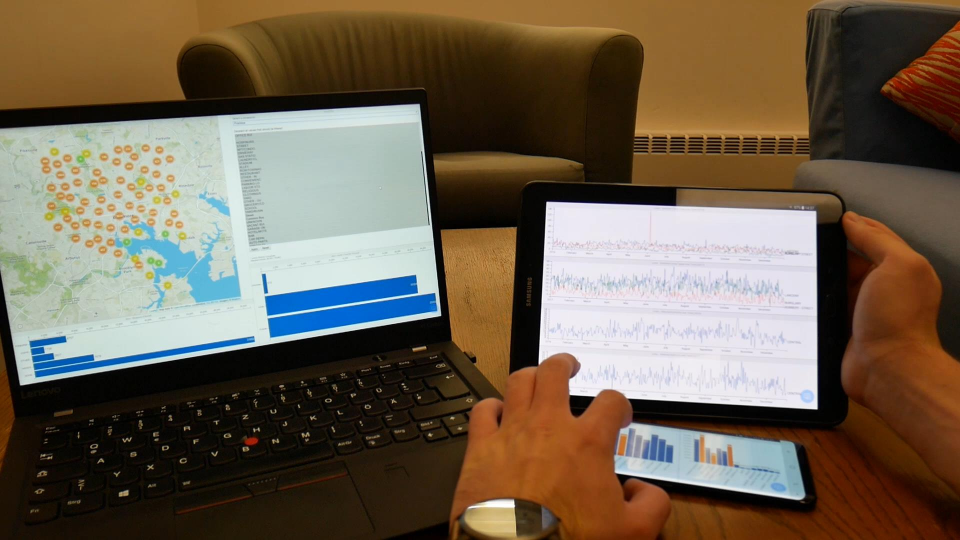}
    }
    \subfloat[Visualization relationships used to guide automatic layout.]{
        \includegraphics[height=4.5cm]{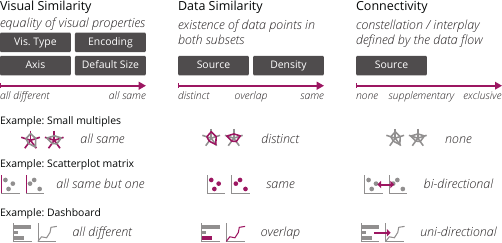}
        \label{fig:vistribute-relation}
    }
    \caption{\textbf{Automatic visualization layout using Vistribute.} 
        The Vistribute system will dynamically reorganize the visualizations to be displayed on the available surfaces whenever the device configuration changes (left).
        Relationships between the different visualizations are used to ensure that similar representations are linked (right).
        For example, the time-series charts on the tablet in the left picture are always laid out vertically on the same display and with a common horizontal axis.}
    \label{fig:vistribute}
\end{figure*}

One example where such automated resource management is critical is display management. 
In a typical ubiquitous analytics scenario, we might imagine that the displays available to a person will change dynamically over the course of a work session.
For example, while leaving your car or bus on the way to the office, you might only have access to your smartphone, allowing you to view a single data display at a time.
Once you get to your office and your desk, you can boot up your computer and distribute an entire visualization dashboard across your monitors as well as your personal devices.
Finally, when you head into the conference room for a meeting, the projected screen as well as the laptops belonging to your colleagues could be used to display even more detailed data visualizations.
However, for this to be practical, the layout of visualizations across available displays should be automatic lest the user gets continually bogged down in moving charts around.
Furthermore, such layout must organize closely related charts together on the same display as well as respect their geometry affinity.

Such dynamic layout across ubiquitous analytics environments was our focus in the Vistribute~\cite{DBLP:conf/chi/HorakMKDE19} system (Figure~\ref{fig:vistribute}) from 2019.
Based on a design space of visualization relationships (Figure~\ref{fig:vistribute-relation}), the Vistribute layout manager performs real-time constraint solving based on a set of high-level visualization layout heuristics.
These heuristics include multi-chart relationships, such as the visual and data similarity as well as data flow connectivity in the figure, but also single-chart heuristics.
For example, dense data displays, such as geographical maps, are given more display space, and a skewed-aspect ratio chart, like a time-series line chart, will be given wide and short display allocations.

The VisHive toolkit~\cite{Cui2018} tackles a different but similar problem: computational resource management. 
More specifically, mobile ubiquitous analytics often call for significant computation, such as when performing textual analysis, clustering, or machine learning. 
The computational power of an individual mobile device may be insufficient to complete this calculation on its own in a timely manner.
One solution is to simply use a remote computational resource, but even if the internet connectivity would be there---which is not a given in many mobile on-the-go situations---the data to be processed is often simply too large to effectively upload on the fly.
The VisHive system solves this by forming an ad-hoc cluster, or \textit{hive}, of mobile devices for load balancing computation in the field.
Designed to work in conjunction with web-based visualization systems, VisHive is a small JavaScript library that runs directly in a device's web browser and requires no specialized software to be installed on the device.
In this way, a user can simply bring additional mobile devices online to help process computational tasks in case a particular computation is taking too long to complete.

\begin{mdframed}[backgroundcolor=black!5, linewidth=0pt]
\textbf{Takeaway:} Resource management across multiple devices involved in a ubiquitous analytics environment should be automated to minimize the user burden.
\end{mdframed}

\subsection{Multiple Collaborators}

Finally, the third cardinality factor in a ubiquitous analytics system is the individual users that often come together to work on realistic tasks. 
In general, collaborative visualization is a grand challenge of data visualization research, but ubiquitous analytics have the benefit of being designed to be collaborative from the very foundation.
For example, the David and Goliath system discussed above can easily be used in a collaborative manner (as in Figure~\ref{fig:david-and-goliath}), and the two different interfaces in ReLive could just as easily be employed by two parallel users rather than the same one in sequence.

\begin{figure}[htb]
    \centering
    \includegraphics[width=\linewidth]{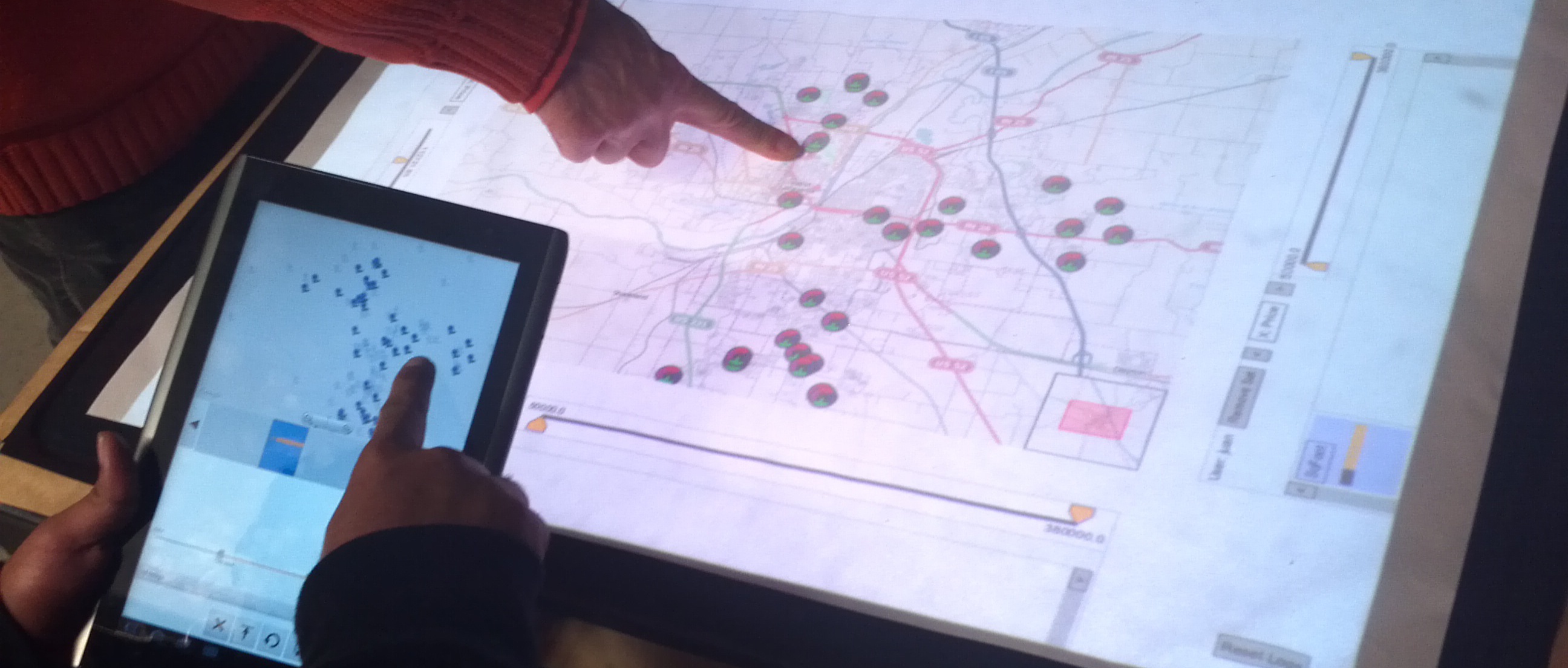}
    \caption{\textbf{Collaboration using Branch-Explore-Merge.} 
        The mobile device, one per user, is generally synchronized with the data display on the tabletop.
        However, if the user wants to deviate from the shared display, they can branch the current state of the tabletop on their personal device, make the desired changes, and then merge back their findings to the shared display.
        Merging requires consensus.
    }
    \label{fig:bem}
\end{figure}

One example of this is the Branch-Explore-Merge tabletop system~\cite{McGrath2012} that we proposed in 2012 for collaboration across tabletops and mobile devices (Figure~\ref{fig:bem}).
A precursor to the D\&G system discussed earlier, Branch-Explore-Merge uses the same philosophy of private and personal devices (smartphones and tablets) vs.\ public and shared devices (large touch tabletops in this case) as in the smartwatch and wall display scenario.
However, with B-E-M the focus is specifically on the coordination and consensus mechanisms required for a visualization system used by multiple users.
While a user should be able to modify the visualization on their own personal device at will, any change made on the shared tabletop will directly affect all collaborators.
To manage this process, B-E-M draws on basic revision control principles from software engineering, where the user can branch the current state of the shared display on their own device, explore the data on their own, and then finally merge back their changes---or discard them if the exploration turns out to be a dead end.
The B-E-M system requires a vote from all participants around the tabletop for changes to be merged back onto the shared display.

\begin{figure}[tbh]
    \centering
    \subfloat[Comparing stacked lenses.]{
        \includegraphics[width=0.48\linewidth]{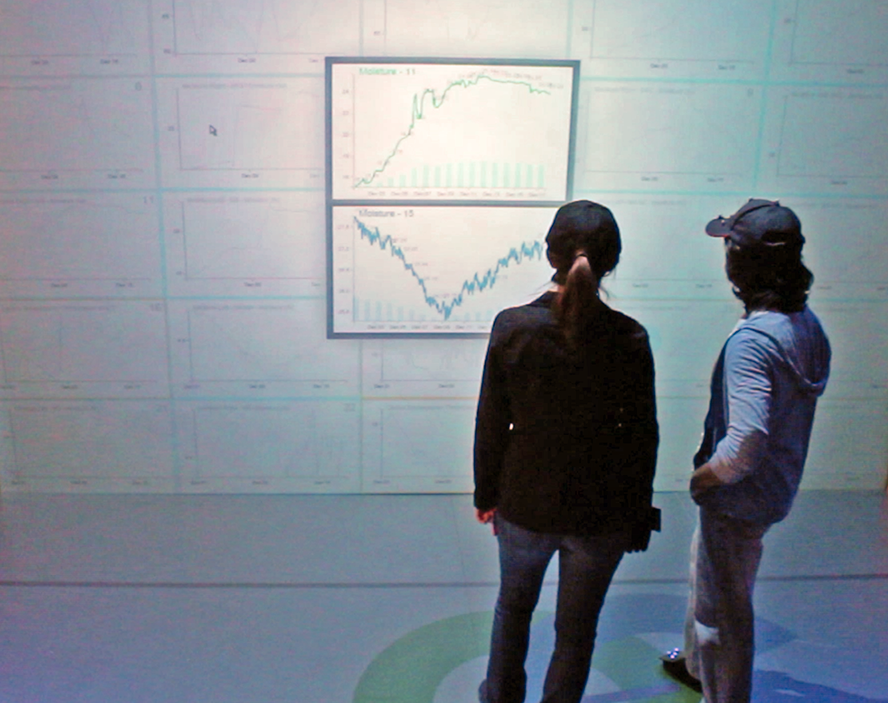}
    }
    \subfloat[Consensus to overlay lenses.]{
        \includegraphics[width=0.48\linewidth]{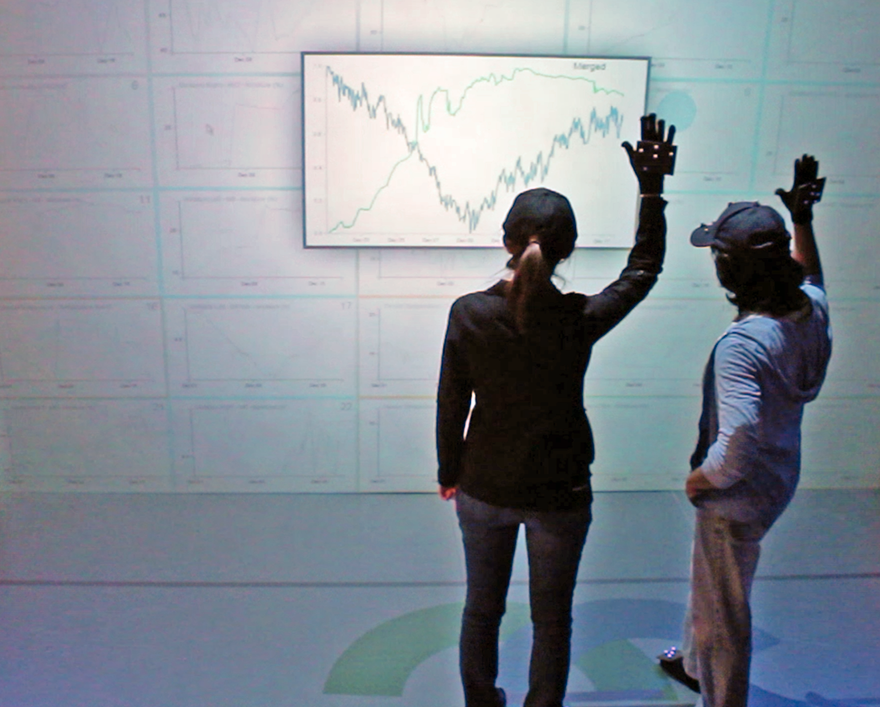}
        \label{fig:proxemic-consensus}
    }
    \caption{\textbf{Collaborative data analysis using the Proxemic Lens.}
        In the left image, the two collaborators have approached each other sufficiently that their body language indicates their work is closely coupled, so their respective lenses are stacked vertically.
        In the right image, the collaborators both raise their hands in consensus, causing the lenses to be overlaid to allow for direct comparison in the same visual space.
    }
    \label{fig:proxemic-lens}
\end{figure}

Voting is a somewhat disruptive coordination mechanism, and often more implicit mechanisms are better.
In the Proxemic Lens~\cite{Badam2016b} project from 2016, we used proxemics information~\cite{Greenberg2011} such as the distance between users, the direction of their bodies and head, their hand and foot gestures, and the spatial arrangement of objects and people in 3D space to guide interaction.
The goal of the project was to be able to infer user intention in a ubiquitous analytics scenario from their body language and physical navigation in a space.
We were particularly interested in studying how a large shared display can be best utilized depending on the collaborative coupling of the people using it; separate viewports (lenses) for people working independently, and stacked or overlaid viewports for people working closely together (Figure~\ref{fig:proxemic-lens}).
We used a static environment and a floor-to-ceiling display with a Vicon motion capture system to track this information for multiple collaborators, but in a mobile environment more subtle biometric or inertial sensing technology will be required.
We found that our participants overall enjoyed the implicit interaction of the Proxemic Lens system and that their intention was often inferred correctly. 
However, they also indicated a preference for explicit rather than implicit gestures for actions that are typically seen as commands, such as creating charts, splitting viewports, and consensus operations (Figure~\ref{fig:proxemic-consensus}).

\begin{mdframed}[backgroundcolor=black!5, linewidth=0pt]
\textbf{Takeaway:} Collaboration in ubiquitous analytics requires careful consideration of coordination and consensus, just like for general collaborative work, but the heterogeneous devices typically employed in ubiquitous analytics settings makes such coordination mechanisms straightforward to integrate.
\end{mdframed}

\section{Challenges \& Outlook}

This future vision for data analytics is still a new notion, and its various manifestations as visual, ubiquitous, immersive, and situated analytics are still nascent and emergent. 
My research group has been a significant driver in this field, but the story is much bigger than just our efforts.
The greater research community is energetic and growing, with new and exciting analytical systems being proposed at every major conference and journal issue; certainly too many to discuss in a single review article.
However, we claim that the techniques and technologies described here are a representative cross-section of ubiquitous analytics research. 

Based on this review, we would summarize our takeaways using a single theme: \textbf{device diversity}.
Basically, it is the varied and heterogeneous nature---as well as their effective utilization---of the individual devices involved in a sensemaking task that makes ubiquitous analytics powerful.
This idea is also supported by the post-cognitive frameworks discussed earlier in this article.

If heterogeneity is the lead theme of ubiquitous analytics, where is the field going in the future?
Ens et al.~\cite{Ens2021} presented a vision for immersive analytics in 2021 and outlined the grand research challenges of that field. 
We are supportive of these challenges in general, but we here complement their technical nature with our own list of higher-level future challenges:

\vspace{-0.15cm}

\begin{itemize}

    \item\textbf{The Future is Mixed:} While handheld devices with screens are here to stay, it seems almost inevitable that future devices will eventually be based on Augmented and Mixed Reality~\cite{speicher19whatismr}.
    AR/MR displays have the capacity to turn your entire surrounding world into a canvas for visual data displays, and the technology is constantly improving.
    This should mean that immersive 3D visualizations and interaction techniques will become increasingly important in the future.
    
    \item\textbf{Human-Centered Artificial Intelligence (HCAI) Teaming:} While mostly glossed over in this article, the future of ubiquitous analytics is closely entwined with artificial intelligence.
    Only with powerful automated algorithms and recommendations at our beck and call will we be able to overcome the computational challenges of tomorrow.
    However, rather than the black-box pipeline model of traditional AI, we believe in the use of visual interactive interfaces as inflection points for involving human operators in the loop.
    This is known as \textit{human-centered artificial intelligence} (HCAI)~\cite{Shneiderman2022}, but this is outside the scope of this article.
    
    \item\textbf{Standards, Components \& Practices:} If future sensemaking environments are characterized by heterogeneous systems and devices, then we will need standards as well as standardized components and practices to enable multiple vendors to provide their own versions of these tools.
    While there exist several immersive analytics~\cite{cordeil19iatk} and Augmented Reality toolkits~\cite{Fleck2022}, we will need a much richer ecosystem, including for software development, session and view management, navigation, etc.
    Furthermore, many existing toolkits are built on proprietary 3D game engines such as Unity.
    It is difficult to predict the future here, but based on past history, technologies based on open standards, such as the web-based VRIA~\cite{butcher20vria} and Vistrates~\cite{Badam2019} toolkits, are safer bets than those built on closed and proprietary technology.

    \item\textbf{Accessibility \& Inclusivity:} The field of data visualization is just now recognizing the accessibility of interactive visual representations to people with visual, motor, or cognitive impairments, etc.
    Ubiquitous analytics will need to learn from these lessons at an early stage. 
    However, the heterogeneous nature of ubiquitous analytics can once again play in our favor in that the inclusion of diverse platforms into the sensemaking loop may make it easier to accommodate people of varying physical and cognitive ability.

    \item\textbf{Of Scale and Scalability:} With the exception of the VisHive project, most of this paper has not concerned itself with large scale data management.
    While data visualization is often seen as a best-in-class solution for the human aspect of big data~\cite{DBLP:journals/interactions/FisherDCD12}, there are many challenges on the computational and data management side that must be solved in other to enable mobile sensemaking at this scale and magnitude.

    \item\textbf{Evermore Everywhere:} Similarly, while the everywhere data aspect of this paper is more of about providing opportunity for anywhere data, we still need further work on capturing, integrating, and synthesizing heterogeneous data from multiple sources in our environment.
    While privacy and security must remain foremost in our minds, such new data about our world will only continue to make data-driven decision-making better and more effective.
    
\end{itemize}

The current state of ubiquitous data analytics has been more than ten years in the making.
There is now a thriving and creative research community that is invested in taking this vision forward into the future.
Perhaps one day we will truly see sensemaking become embedded into the fabric of everyday life.